\documentclass[journal]{IEEEtran}
\IEEEoverridecommandlockouts
\usepackage{mathrsfs}
\usepackage{amsfonts}
\usepackage{ifpdf}
\usepackage{cite}

\usepackage{xcolor}

\ifCLASSINFOpdf
\usepackage[pdftex]{graphicx}
\else \fi
\usepackage[cmex10]{amsmath}
\usepackage{array}
\usepackage{mdwmath}
\usepackage{mdwtab}
\usepackage{eqparbox}
\usepackage[tight,footnotesize]{subfigure}
\usepackage{algorithmic}
\usepackage{algorithm}
\usepackage{amsmath}
\usepackage{epsfig}
\usepackage{ae}
\usepackage{amssymb}
\usepackage{times}
\usepackage{amsthm}
\usepackage{amsmath}   

\begin{document}

\title{Ultra-Dense HetNets Meet Big Data: Green Frameworks, Techniques, and Approaches}

\author{
         Yuzhou~Li,
         Yu~Zhang,
         Kai~Luo,
         Tao~Jiang,
         Zan~Li,
         and Wei~Peng
\thanks{Y. Li, Y. Zhang, K. Luo, T. Jiang, and W. Peng are with Huazhong University of Science and Technology and Z. Li is with Xidian University.}
}
\maketitle
\IEEEpeerreviewmaketitle
\begin{abstract}
Ultra-dense heterogeneous networks (Ud-HetNets) have been put forward to improve the network capacity for next-generation wireless networks. However, counter to the 5G vision, ultra-dense deployment of networks would significantly increase energy consumption and thus decrease network energy efficiency suffering from the conventional worst-case network design philosophy. This problem becomes particularly severe when Ud-HetNets meet big data because of the traditional reactive request-transmit service mode. In view of these, this article first develops a big-data-aware artificial intelligent based framework for energy-efficient operations of Ud-HetNets. Based on the framework, we then identify four promising techniques, namely big data analysis, adaptive base station operation, proactive caching, and interference-aware resource allocation, to reduce energy cost on both large and small scales. We further develop a load-aware stochastic optimization approach to show the potential of our proposed framework and techniques in energy conservation. In a nutshell, we devote to constructing green Ud-HetNets of big data with the abilities of learning and inferring by improving the flexibility of control from worst-case to adaptive design and shifting the manner of services from reactive to proactive modes.

\end{abstract}
\section{Introduction}
\subsection{Ultra-Dense Heterogeneous Networks}
With the flourish of Internet of Things (IoT), the amount of mobile devices increases explosively and correspondingly the massive connections between humans, humans and machines, and machines skyrocket continuously. These in turn boost the exponential growth of wireless traffic volumes, resulting in 1000-fold data challenge to wireless networks  \cite{GreenHCRAN_TechniquesTradeoffChallenges_ICM2017,add1_UDN}.
On the other hand, available spectrum resources that can be allocated to wireless networks become scarcer and scarcer, it is thus
impractical to dramatically improve the network capacity by continuously obtaining additional spectrums.
As a result, more advanced network-organization and wireless technologies are eagerly needed to improve the network capacity.

One of the most promising approaches to solve the aforementioned challenges is to substantially densify the deployment of small cells overlaying the conventional macro cells\cite{JointOptimization_HetNets_JSAC2016}.
Through reusing the limited spectrums among cells, the network spectrum utilization and thus the network capacity can be dramatically improved.
This kind of network-organization architecture is referred to as ultra-dense heterogeneous networks (Ud-HetNets), and has been recognized as a key technology for 5G wireless communications \cite{5G_VisionAndRequirements}. In Ud-HetNets, multiple radio access networks (RANs), e.g., UMTS, 4G, and WiFi, coexist, and traditional high-power macro cells and low-power small cells such as pico-cell, femtocell, and relay \cite{add2_UDN}, operating in either licensed or unlicensed spectrums, are densely mixed. By abundantly reusing spectrums across cells and RANs and bringing antennas close to users, the data rate of users can be increased and their energy consumption can also be reduced.

\subsection{Ud-HetNets Meet Big Data}
Ud-HetNets are bound to networks of big data. Firstly, a large number of base stations (BSs) with different radio technologies are deployed in networks, which undoubtedly produce a great quantity of control data. Secondly, tremendous smart devices with diversified information and records, such as social relations and request histories, also generate a large amount of data. Last but not least, various kinds of communications and interactions, e.g., image, video, and multimedia, further skyrocket the network data.

Ud-HetNets meeting big data, i.e., Ud-HetNets of big data, impose lots of challenges on their ubiquitously practical deployment, especially in the aspect of the incurred huge energy consumption. Specifically,
\begin{itemize}
  \item \textbf{\textit{Big network interference}}. The universal spectrum reuse across cells and RANs incurs severe inter-cell and cross-network interference, which significantly degrade the network energy efficiency without effective interference coordination schemes.
  \item \textbf{\textit{Big small-cell deployment}}. More energy expenditure is possibly required to operate densely deployed small cells compared to traditional macro-cell-only networks.
  \item \textbf{\textit{Big transmit energy}}. Energy consumed to deliver the big network data will be quite considerable. In addition, the traditionally reactive service mode that delivers data only after receiving requests also degrades energy efficiency.
  \item \textbf{\textit{Big static energy waste}}. It is rather energy-expensive to always switch on all BSs as in the widely-adopted worst-case network planning approach, because mobile traffic loads usually vary in both spatial and temporal domains  in Ud-HetNets.
\end{itemize}

However, it is explicitly expected in 5G that the network energy efficiency, compared with 4G,  needs to be improved by more than $100$  times \cite{5G_VisionAndRequirements}. Therefore, energy-efficient frameworks, techniques, and approaches for Ud-HetNets of big data should be developed.

\subsection{Energy-Efficient Design for Ud-HetNets of Big Data}
To energy-efficiently deploy Ud-HetNets, it is required to improve energy efficiency degraded by inter-cell interference, to reduce the additional energy consumption incurred by the traditional request-transmit mode, and to avoid huge energy waste from always switching on all BSs. For these targets, \textit{Ud-HetNets of big data should discern network patterns and perceive user behaviors to make adaptive and proactive decisions for energy-efficient network control, optimization, and management.} This requires that the network paradigm for Ud-HetNets of big data should be equipped with intelligence.

Luckily, recent breakthroughs in artificial intelligence (AI) especially in machine learning make proactive and intelligent network design possible. Specifically, characteristics hiding in the generated big data, such as network patterns and user behaviors, can be mined and predicted with the help of AI \cite{BigDataSurvey_MNA2014}. Then combined with learning techniques, those extracted information can be used to establish energy-efficient decision making systems. Moreover, benefiting from the significant developments in low-complexity algorithm design in machine learning, cost-efficient algorithms for energy conservation can also be developed for practical applications in Ud-HetNets.

The remainder of this article is organized as follows. In Section~\ref{Section:Architecture_UdHetNets}, we develop a big-data-aware AI-based framework for energy-efficient operation of Ud-HetNets. Based on the framework, Section~\ref{Section:ImportantResearchDirections} introduces and analyzes four promising energy-efficient techniques. In Section~\ref{Section:LoadAwareEnergyEfficientOptimizationModel}, we develop a load-aware energy-efficient optimization model to show the potential of our proposed framework and techniques in energy conservation. Finally, we conclude our paper in Section~\ref{Section:Conlusions}.

\section{A Big-Data-Aware AI-Based Network Framework} \label{Section:Architecture_UdHetNets}
In this section, we first introduce the framework design guidelines for energy-efficient operations of Ud-HetNets, and then explain the main components of the proposed framework.

\subsection{Overall Design Guidelines}

\textbf{\textit{Change the mode of services}}. The traditional request-transmit-consume mode is energy-expensive for Ud-HetNets of big data, where data is delivered to users only when their requests are received by networks.
In fact, the popularity of contents and users' typical behaviors, e.g., preference, influence, and mobility, which seem random but regular intrinsically, can be (approximately) precisely predicted. Hence, some of contents, originally stored at network servers far from users and delivered to users only when receiving requests, can be pre-pushed to the network entities (e.g., BSs) near users even to the user sides during good network conditions periods (e.g., small interference). \textit{This is a change of network service philosophy, from reactive request-transmit-consume to proactive transmit-request-consume modes.} By this, the energy efficiency of Ud-HetNets of big data can be greatly improved.

\textbf{\textit{Improve the flexibility of control}}. The traditional worst-case network design philosophy devotes to satisfying users' demands even at peak traffic hours.
This inevitably leads to a large amount of energy waste and a lot of network energy efficiency reduction due to temporally and spatially variant traffic loads in realistic networks.
For example, the time fraction when the traffic is below 10\% of the peak during a day is about 30\% and 45\% in weekdays and weekends, respectively \cite{DynamicTrafficLoads_CellularNetwork}. Therefore, a large number of BSs are extremely underutilized during off-peak periods in cells-densely-deployed Ud-HetNets. However, BSs still consume more than 90\% of their overall power consumption even with little and no activity, e.g., a typical UMTS BS consumes 800--1500W but only with RF output power of 20--40W \cite{Energy_Delay_Association_JSAC2011}. \textit{It is therefore important to improve the flexibility of network control by shifting from the worst-case to adaptive operations for energy conservation.}

\subsection{Framework Design}
\begin{figure}[t]
\centering \leavevmode \epsfxsize=3.5in  \epsfbox{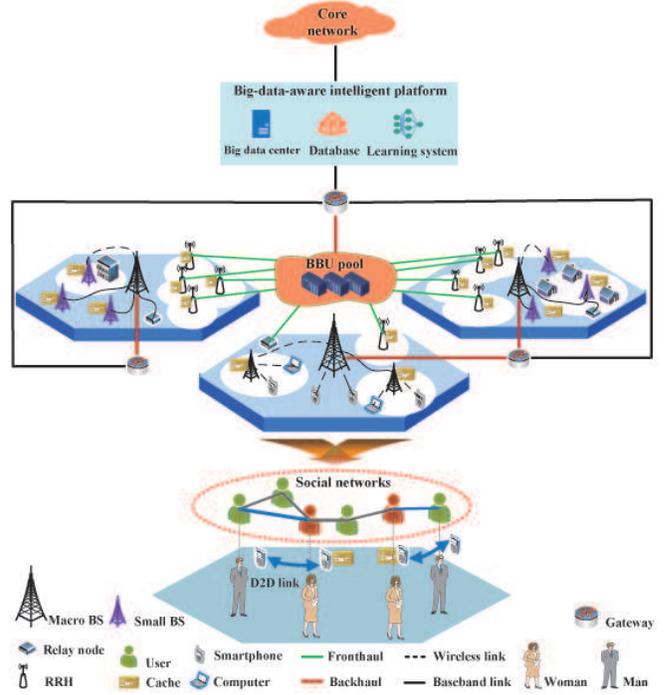}
\centering \caption{Big-data-aware AI-based network framework for energy-efficient Ud-HetNets.} \label{Fig:Intelligent_EnergyEfficient_Framework}
\end{figure}

Following our design guidelines above, we put forward a big-data-aware AI-based network framework, as shown in Fig.~\ref{Fig:Intelligent_EnergyEfficient_Framework}, to energy-efficiently operate ultra-dense HetNets. The main components of the proposed framework are as follows.

\textbf{1) Centralized BBU pool}

In Ud-HetNets, macro BSs (MBSs) are kept for improving mobility performance and reducing Ping-Pong effects while parts of small-cell BSs (SBSs) are split into baseband units (BBUs) and remote radio heads (RRHs), with their interconnection via high-bandwidth low-latency optical transport networks.
Then BBUs are integrated into centralized BBU pools, responsible for intelligent resource allocation and signal processing with the help of cloud computing and virtualization techniques while RRHs for information radiation.
MBSs and BBU Pool are connected to the big-data-aware intelligent platform through gateways, which is then associated with the core network.
Thus, the centralized BBU pools with strong computing and coordinating abilities facilitate the sharing of information among cells and RANs, which paves the path for intelligent resource optimization.

\textbf{2) Social networks}

With the ever-increasing popularity of social softwares, such as Facebook and Wechat, social networks have become an important part of Ud-HetNets of big data.
Unlike communication entities (e.g., MBSs and BBUs), social networks, almost irrelevant to physical entities, are used to describe the relationships between/among users in networks, regardless whether they are close to each other or located in different areas.
For example, if a person is in the circles of friends of another person in Wechat, then it is quite possible for them to visit and share the same contents.
To boost the energy efficiency of Ud-HetNets, we can exploit the features of social networks such as popularity of contents and users' social relationships. For example, the systems can pre-cache popular contents in influential users' smart devices and share these contents via social networks in the way of device-to-device (D2D) communications when they are close to each other \cite{AdmissionControlResourceAllocation_D2DNetworks_JSAC2016,add3_cache}. By allowing D2D communications and pre-caching the contents, users' transmit power can be significantly decreased attributed to the reduction of transmit distance and the avoidance of traffic congestion.

\textbf{3) Big-data-aware intelligent platform}

\begin{figure}[t]
\centering \leavevmode \epsfxsize=3.5in  \epsfbox{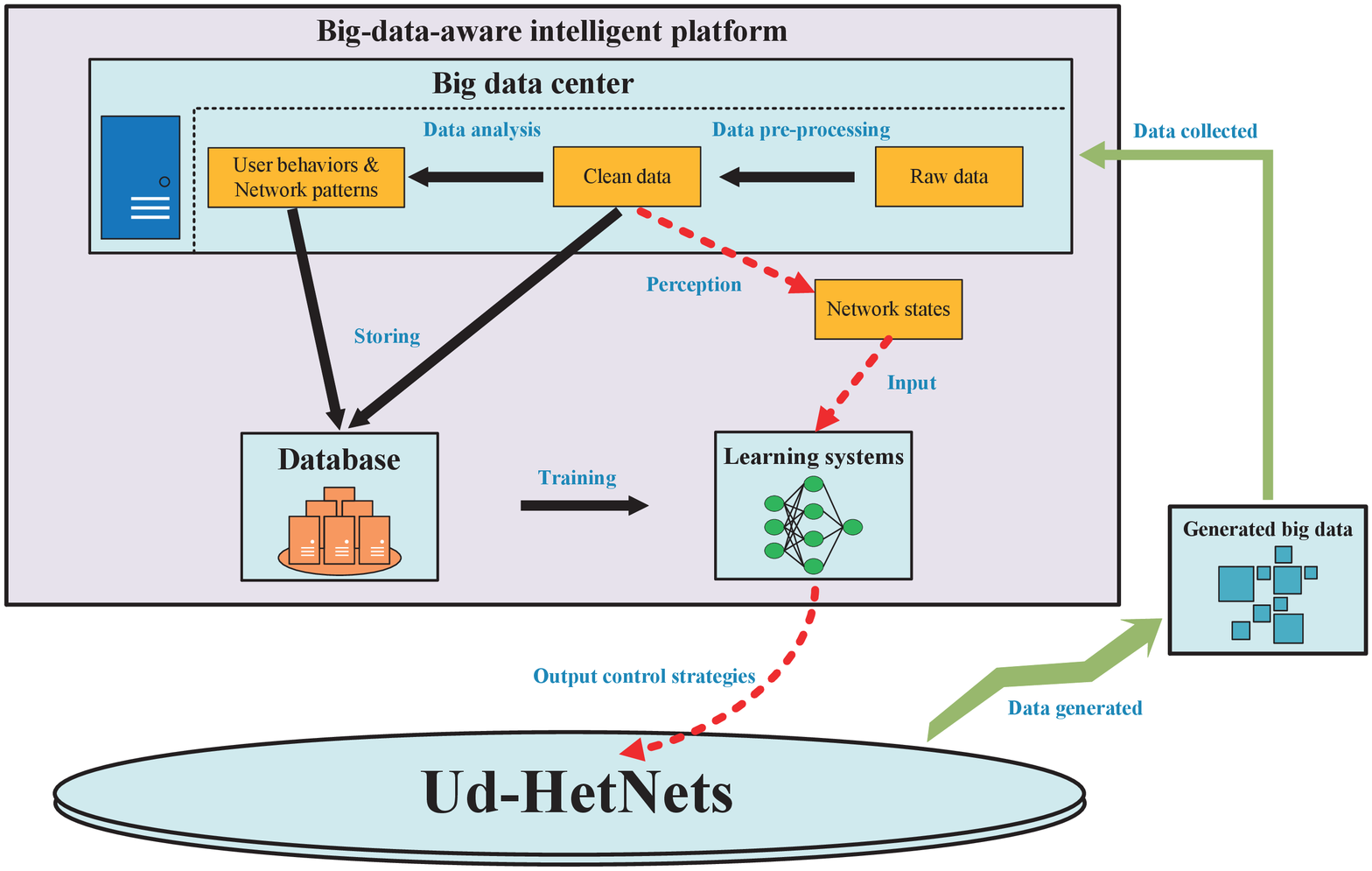}
\centering \caption{Data analysis and learning procedures on the big-data-aware intelligent platform.} \label{Fig:Process_of_Data_Analysis_and_Learning}
\end{figure}

As shown in Fig.~\ref{Fig:Intelligent_EnergyEfficient_Framework}, the big-data-aware intelligent platform is implemented between the gateway and the core network, consisting of a big data center, a database, and a learning system. Fig.~\ref{Fig:Process_of_Data_Analysis_and_Learning} further overviews the process for data analysis and learning on the platform. First, raw data generated in Ud-HetNets is collected and pre-processed by the big data center. Then, through AI-based data analysis, user behaviors and network patterns are extracted from clear data and related results as well as useful history data are stored in the database, which are later used to train the learning system. To make intelligent reactions, the big data center also perceives the real-time network states, e.g., user distribution. These states are further treated as the input of the learning system, which finally outputs intelligent strategies for network control, optimization, and management.

\section{Potential Research Directions and Possible Challenges} \label{Section:ImportantResearchDirections}

In this section, we identify four promising research directions for energy conservation in the proposed framework and illustrate the possible challenges when solving these issues.

\subsection{Big Data Analysis}
\begin{figure}[t]
\centering \leavevmode \epsfxsize=3.5in  \epsfbox{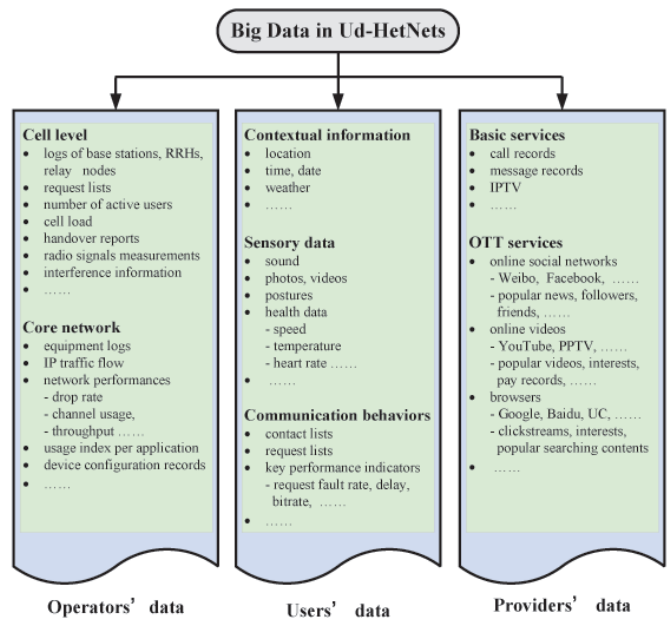}
\centering \caption{Classification of the collected big network data in Ud-HetNets.} \label{Fig:BigDataClassification}
\end{figure}

A process of an information interaction (e.g., web-surfing) involves three sides: users, wireless operators, and service providers. Based on this fact, the generated network data in Ud-HetNets roughly falls into three categories, namely users', operators', and providers' data. Fig.~\ref{Fig:BigDataClassification} elucidates some of main components of the generated big data in Ud-HetNets, which are explained as follows.

\begin{itemize}
  \item \textit{Users' data}: Collected from users' devices especially various embedded sensors, it consists of a lot of contextual information and users' daily behaviors, such as locations and request lists.
  \item \textit{Operators' data}: It is mainly from densified cells and core networks, which produce abundant cell-level information (such as cell loads) and network-layer data (e.g., network performance measurements), respectively.
  \item \textit{Providers' data}: The over-the-top (OTT) services such as online video applications and browsers are the main source. Those applications can record users' visited contents and rank content popularity, from which personal preferences and group information can be mined.
\end{itemize}

After collecting and pre-processing these massive raw data, we need to refine useful information hidden in them. \textit{For energy saving, at least two kinds of information, namely network patterns and user behaviors, are needed to mined out based on AI for our framework}.

\textbf{\textcircled{1} Network patterns}: By global information sharing, the centralized BBU pool proposed in the framework facilitates overall energy-efficient network design across cells and RANs. For this, we need to extract traffic information from the large volume of traffic data at network sides.
\begin{itemize}
  \item \textit{Content popularity}: It describes the popularity of on-line contents. Some contents are more popular and possibly become popular in the near future in certain small cells or in overall networks. If those contents are pre-delivered and locally cached, repeated transmissions can be avoided, and thus transmit energy can be reduced.
  \item \textit{Temporal variations}: It describes the variation rules of traffic volumes with time. For a given area in Ud-HetNets, its traffic volumes usually vary but does not fluctuate sharply  with time.
  \item \textit{Spatial variations}: It describes the variation rules of traffic volumes with space. For a given time, traffic volumes usually vary in different areas. Note that spatial-temporal variations of traffic loads usually resemble day by day, which are valuable for future traffic distribution prediction and adaptive network planning.
\end{itemize}

\textbf{\textcircled{2} User behaviors}: Recent results showed that human behaviors, such as users' web-surfing customs, are correlated and predictable to a large extent \cite{LimitsOfPredictability_HumanMobility}. Users' individual behaviors can be exploited to guide contents distribution across social communities and determine the pre-pushed contents for individual users. In our proposed framework, the following information needs to be mined from the obtained historical data for the subsequent energy saving optimization.
\begin{itemize}
  \item \textit{Personal preference}: It reflects users' web-surfing customs and preference on on-line contents. Obtaining users' performance is helpful to recommend and pre-push appropriate contents for them.
  \item \textit{Social influence}: It reflects the degree of users' activity and their influence. Influential users are persons who are popular in communities or among certain groups. That is, the online contents shared by them are usually clicked by others with higher possibilities.
  \item \textit{Mobility regulation}: It reflects users' daily mobility customs, such as trajectory, daily routine, and frequent places. It is helpful to make predictions of users' next cells for seamless handovers and pre-allocate resources for them to ensure QoS requirements.
\end{itemize}

\subsection{Adaptive BS Operation}
From big data analysis, we have known temporal-spatial traffic variation rules from network patterns.
Hence, it is a waste of energy to keep all the BSs always on as in the traditional worst-case network management.
\textit{The general purpose of adaptive BS operation is to activate an appropriate number of BSs to support current traffic loads and to turn off the remaining BSs for energy saving\footnote{Note that, From Fig.~\ref{Fig:Intelligent_EnergyEfficient_Framework}, our framework includes multiple network transmit devices, such as MBS, SBS, RRH, and relay. For notational simplicity, BS operation denotes switching on/of all these devices throughout the article.}.} By this, network deployment macroscopically matches both spatial and temporal dynamics of traffic loads, and thus energy saving is achieved on large scales by reducing the static power.

There have been some works to address the BS operation problems from different perspectives \cite{CellZoomingCostEffcient_CM2010,GreenDynamicBaseStation_TWC2013,Energy_Delay_Association_JSAC2011,BSOperation_UserAssociation_CL2013}. In \cite{CellZoomingCostEffcient_CM2010}, the authors proposed the technique of cell zooming, where the cell size is tuned with traffic load fluctuations for energy controlling. The notion of network-impact was introduced in \cite{GreenDynamicBaseStation_TWC2013} to investigate the energy saving problem through dynamic BS switching. Further, \cite{Energy_Delay_Association_JSAC2011} and \cite{BSOperation_UserAssociation_CL2013} jointly optimized BS operation and user association to reveal energy-delay and energy-revenue tradeoffs, respectively. However, although \cite{CellZoomingCostEffcient_CM2010,GreenDynamicBaseStation_TWC2013,Energy_Delay_Association_JSAC2011,BSOperation_UserAssociation_CL2013} took  spatially inhomogeneous traffic distributions into account, they did not consider temporal traffic variations, which also dramatically affects network energy consumption.

In addition, the following aspects also impose challenges on BS operation to be applied to realistic wireless networks.
\begin{itemize}
  \item What is an appropriate time scale for BS operation? Because frequent BS on/off operation would cause additional signaling overhead and network oscillation.
  \item How to guarantee that switching BS off for energy saving would not significantly decrease users' QoS, e.g., rate?
  \item How to mathematically formulate and flexibly balance network-user energy cost tradeoffs? This is because switching BSs off for network energy saving would increase users' transmit power due to the increased distance between users and BSs.
\end{itemize}

\subsection{Proactive Caching}
With the development of semiconductor technologies, equipping both network and user devices with a large volume
of storage becomes possible.
Leveraging the derived content popularity and user behaviors resulted from big data analysis, \textit{popular contents of networks and contents possibly visited by individual users can be pre-pushed from network servers to BSs or user equipments (UEs)}. Hence, proactive caching can be classified into two categories: \textit{BS-level and UE-level caching}. Compared with the BS-level caching, UE-level caching further pre-pushes the contents cached at BSs to UEs. At the risk of consuming more energy, UE-level caching is more energy-efficient if the pre-pushed contents do be visited by users. Regarding proactive caching, the following aspects therefore need to be addressed.
\begin{itemize}
  \item How to define the popularity of online contents? In other words, what is the popularity threshold to pre-push a content?
  \item How to determine the pre-pushed contents for individual users according to their behaviors and how many of these contents are pre-cached?
  \item Which, BS-level caching or UE-level caching, should be chosen when starting proactive caching?
\end{itemize}

Although it is attractive to reduce transmit energy via proactive caching, inappropriate contents pre-pushing incurs additional waste of energy and storage, e.g., pre-cached contents are never visited by users. Hence, besides the above issues related to proactive caching itself, we also need to carefully consider the following revenue-cost tradeoff problems:
\begin{itemize}
  \item How many BSs for BS-level caching or UEs for UE-level caching are selected?
  \item When to start the content pre-pushing and pre-caching in advance of users' formal requests?
  \item How to devise robust proactive caching schemes against random and time-variant wireless networks?
  \item How to mathematically formulate revenue-cost tradeoffs and flexibly balance them?
\end{itemize}

These problems are still under exploration currently and no clear results have been obtained so far.

\begin{figure*}[t]
\centering \leavevmode \epsfxsize=6in  \epsfbox{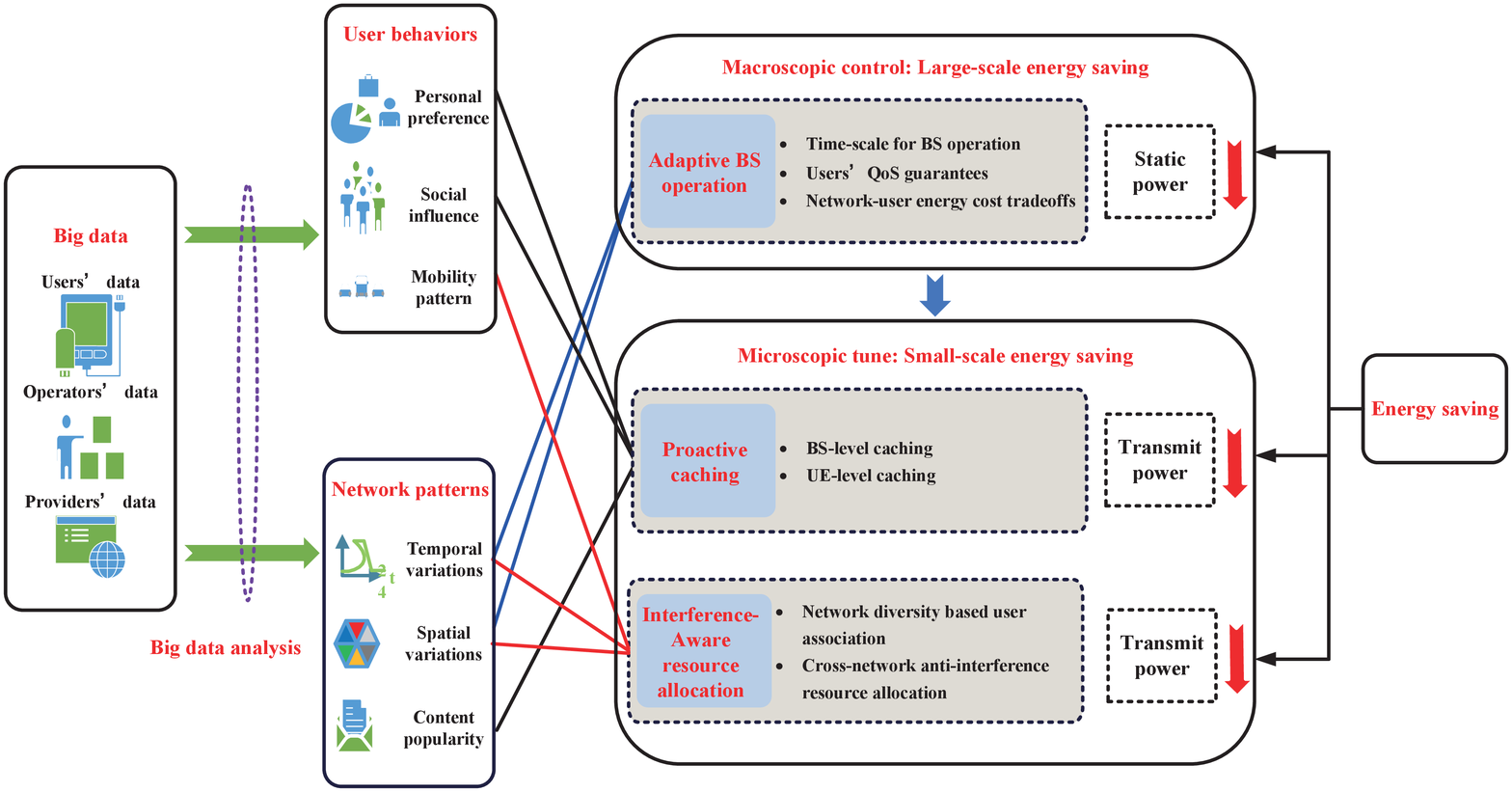}
\centering \caption{The relations among the four promising techniques and their contributions on energy saving.} \label{Fig:ContentsOutlines}
\end{figure*}

\subsection{Interference-Aware Resource Allocation}

Although the amount of transmit power from resource allocation compared to the total operational power is small, it exerts substantial influences on the required power for amplifiers and cooling systems. For example, \cite{EEImprovementMicroSites_Globecom2009} showed that a MBS can reduce the total power consumption from 766 W to 532 W just by reducing its transmit power from 20 W to 10 W. Hence, \textit{besides reducing static power through BS operation on large scales in Ud-HetNets, it is also indispensable to decrease transmit power on small scales leveraging resource allocation by exploiting the big data analysis results.}


\textbf{Network diversity based user association}: \textit{Accompanied with BS operation, user association is another inseparable issue. This is because, once some BSs are turned off, the UEs originally communicating with them need to be re-associated with other BSs.} In Ud-HetNets, multi-RANs and multiple access points (APs) with different coverage ranges and functions coexist. As a result, unlike traditional single-mode UEs communicating only through a fixed RAN, multi-mode UEs could send and receive data concurrently through multiple of them. This unique feature indicates a new type of diversity in Ud-HetNets, \textit{network diversity}, parallel to traditional time, frequency, and space diversity. Hence, regarding user association in Ud-HetNets, we need to first decide which RAN/RANs should be selected and then which AP/APs should be used for transmission for the selected RAN/RANs. This is because, different user association schemes would result in differentiated load distributions among RANs and APs, which in turn affects the working states of RANs and cross-network resource optimization, and thus significantly affects network interference and energy efficiency.


\textbf{Cross-network anti-interference resource allocation}: \textit{BS operation and user association have determined the macroscopic working states of Ud-HetNets, it is further needed to, from the microscopic perspective, optimize heterogenous resources across RANs to achieve energy-efficient transmission.} In Ud-HetNets, both network architectures and resources present great heterogeneity, and the universal reuse of temporal-spectral resources leads to severe network interference. Therefore, it is first required to propose methods that unify and characterize heterogeneous resources scattered in RANs in Ud-HetNets. Then, based on network diversity and cross-network cooperation, by integrating inner-cell and inter-cell resources, algorithms and schemes that jointly optimize these unified heterogeneous resources are needed to design. By this, inter-cell and cross-network interference is coordinated, mitigated, or suppressed, and thus the network energy efficiency can be further improved on small scales.

\subsection{A Summary on These Techniques}
In Fig.~\ref{Fig:ContentsOutlines}, we summarize the relations among the four proposed promising techniques. Specifically, based on AI, user behaviors and network patterns are mined and predicted from the big data generated in the Ud-HetNet, which set foundations for adaptive BS operation, proactive caching, and interference-aware resource allocation. First, leveraging the derived network patterns from big data analysis, adaptive BS on/off operation aims to macroscopically match the temporal-spatial variant traffic loads for energy saving on a large scale from the perspective of reducing static power. Second, by exploiting the analyzed content popularity and user behaviors, proactive caching pre-pushes popular contents and contents possibly visited by users from servers to BSs or UEs to avoid additional transmit power incurred by the conventional reactive mode. Third, joint optimization of heterogeneous resources scattered across RANs mitigates inter-cell and cross-network interference, and thus further reduces transmit power on a small scale. \textit{In a nutshell, the proposed framework and techniques attempt to enable energy-efficient Ud-HetNets with the abilities of learning and inferring by shifting the mode of services from reactive to proactive modes and improving the flexibility of control from worst-case to adaptive design.}


\section{An Approach: A Load-Aware Energy-Efficient Optimization Model} \label{Section:LoadAwareEnergyEfficientOptimizationModel}


Assuming that network patterns and user behaviors have been mined from the collected big data, this section illustrates how to formulate spatial-temporal dynamics of traffic loads, BS operation, and resource allocation into a load-aware energy-efficient optimization model.

We take an orthogonal frequency division multiple access (OFDMA) based downlink Ud-HetNet consisting of a set of BSs $\mathcal{B}$ with $N$ subcarriers as an example. Assuming that the network runs in slots with duration normalized to integer units, thus slot $t$  refers to the interval $\left[ {t,t + 1} \right)$, $t \in \left\{ {0,1,2, \cdot  \cdot  \cdot } \right\}$. We denote the region served by all the BSs by $\mathcal{K} \subset \mathbb{R}^2$ and let $x \in \mathcal{K}$, $i \in \mathcal{B}$, and $\mathcal{B}_{\text{on}}\left(t\right)$ represent a location, BS $i$, and the set of the switched-on BSs in slot $t$, respectively.

To describe spatially and temporally varying traffic loads, we first introduce a random process $\boldsymbol A\left( t \right) = \left( {A\left( x, t \right)} \right)$, where ${A}\left( x, t \right)$ is the amount of new data that arrives to the UE located at $x$ (simplified as UE $x$ below) in slot $t$. From AI-based big data analysis, the statistical information of $\boldsymbol A\left( t \right)$, e.g., distribution and mean, can be obtained. For simplicity, we assume that $\boldsymbol A\left( t \right)$ is independently and identically distributed (i.i.d.) over slots with arrival rate $\boldsymbol \lambda \left( t \right) = \left( {\lambda\left( x, t \right)} \right)$, i.e., $\mathbb{E}\left\{ {\boldsymbol A\left( t \right)} \right\} = \boldsymbol \lambda \left(t\right)$. Hence, $\boldsymbol \lambda\left( t \right)$ can reflect the spatial and temporal variability of traffic loads by $x$ and $t$.

In what follows, we explain how to depict related parameters/indexes in this spatially and temporally varying system. Denote the transmit power and the channel gain from BS $i$ to UE $x$ on subcarrier $n$ in slot $t$ by $P^{n}_{i}\left(x, t\right)$ and $g^{n}_{i}\left(x,t\right)$, respectively. The signal-to-interference-plus-noise-ratio (SINR) at BS $i$ for UE $x$ on subcarrier $n$ in slot $t$ can be computed as $\gamma ^{n}_{i}\!\left(x,t\right) \!=\! \frac{P^{n}_{i}\!\left(x,t\right) g^{n}_{i}\!\left(x,t\right)}{\sum\limits_{j \in \mathcal{B}_{\text{on}}\!\left(t\right), j \neq i} \!\! \int\nolimits_{\mathcal{K}}\!\rho^{n}_{j}\!\left(y,t\right) \! P^{n}_{j}\!\left(y,t\right) g^{n}_{j}\!\left(x,t\right)dy  + {\sigma}^2\!\left(x,t\right)}$, where $\rho^{n}_{i}\left(x,t\right)$ is an indicator variable that is 1 if subcarrier $n$ is allocated to UE $x$ by BS $i$ and 0 otherwise, and $\sigma^2\left(x,t\right)$ is the noise power. Accordingly, the transmit rate on subcarrier $n$ can be computed as $R^{n}_{i}\left(x,t\right) = \log_{2}\left(1 + \gamma ^{n}_{i}\left(x,t\right)\right)$. Let the user association variable $u_{i}\left(x,t\right) = 1$ if UE $x$ in slot $t$ associated with BS $i$ and $u_{i}\left(x,t\right) = 0$ otherwise. Then the rate obtained for UE $x$ from BS $i$ is $ R_i\left(x,t\right) = u_{i}\left(x,t\right){\sum_{n = 1}^N  \rho^{n}_{i}\left(x,t\right)R^{n}_{i}\left(x,t\right)},\forall i,x,t $ and its sum rate is $ R\left(x,t\right) = \sum_{i \in \mathcal{B}_{\text{on}} \left( t \right)}{u_{i}\left(x,t\right) \sum_{n = 1}^N \rho^{n}_{i}\left(x,t\right)R^{n}_{i}\left(x,t\right)},\forall x,t$. Besides, the queue length update equation of UE $x$ is given by $ {Q}\left( x,{t + 1} \right) = \max [{Q}\left( x,t \right) - {R}\left( x, t \right),0] + {A}\left( x, t \right)$.

Furthermore, if BS $i$ is turned on, its sum transmit power is $P_{i}\left(t\right) \!=\! \sum_{n = 1}^{N} \int\nolimits_{\mathcal{K}} u_{i}\left(x,t\right) \rho^{n}_{i}\left(x,t\right) P^{n}_{i}\left(x,t\right) dx$. Accordingly, its overall power consumption can be modeled as $\text{PC}_{i}\left(t\right) = s_i\left(t\right)\big[ \xi_i P_{i}\left(t\right) + P_{i}^{c} \big]$, where the BS operation variable $s_i\left(t\right) = 1$ if BS $i$ is switched on in slot $t$ and 0 if switched off, and $\xi_i$ and $P_i^c$ are constants denoting the inefficiency of power amplifiers and the static power at BS $i$, respectively.

Considering spatial-temporal traffic dynamics, the load-aware energy consumption minimization problem in HetNets that enfolds BS operation and resource allocation can be formulated as the following stochastic optimization problem.
\begin{equation}\label{Eq:OriginalGreenBSAssociationResourceFormulation}
\begin{aligned}
\mathop {\min} ~~&\overline {\text{PC}}  = \sum_{i \in \mathcal{B}} {\overline {\text{PC}}_i}\\
\text{s.t.}   ~~&\text{C1:} ~\overline Q \left( x \right) < \infty,\forall x \in \mathcal{K}\\
              ~~&\text{C2:} ~\text{BS operation constraints} \\
              ~~&\text{C3:} ~\text{User association constraints} \\
              ~~&\text{C4:} ~\text{Subcarrier assignment constraints} \\
              ~~&\text{C5:} ~\text{Power budget constraints} \\
              ~~&\text{C6:} ~\text{Other resource management constraints}.
\end{aligned}
\end{equation}
In (\ref{Eq:OriginalGreenBSAssociationResourceFormulation}), the definitions of average queue length $\overline Q \left( x \right)$  and average power consumption $\overline {\text{PC}}_i$ can be found in \cite{JointOptimization_HetNets_JSAC2016} (cf. Eqs. (9) and (10), respectively), and C1 ensures network stability. Other constraints should be specified according to scenarios and applications. As an example of (1), \cite{JointOptimization_HetNets_JSAC2016} considered a simplified HetNet, where it is assumed that all UEs must and at most be associated with one BS, i.e., $\text{C3:}\sum_{i \in \mathcal{B}} u_{i}\left(x,t\right) = 1, \forall x \in \mathcal{K},t$, and any subcarrier at BSs can be allocated at most to one user, i.e., $\text{C4:}\int\nolimits_{\mathcal{K}} \rho^{n}_{i}\left(x,t\right) dx \leq 1, \forall i \in \mathcal{B},n$.

\begin{figure}[t]
\centering \leavevmode \epsfxsize=3.5in  \epsfbox{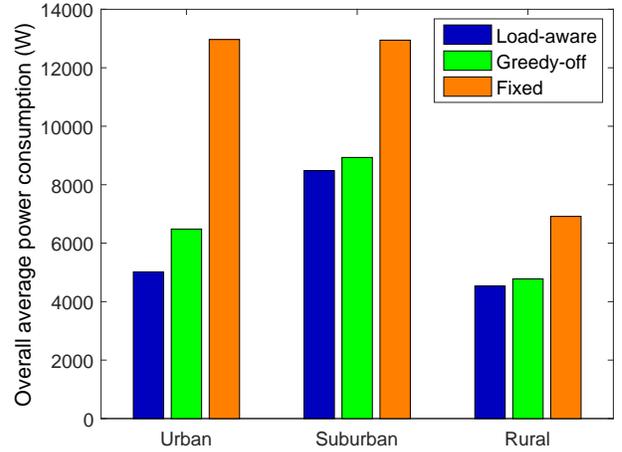}
\centering \caption{Comparison of the overall average power consumption in a real 3G BS deployment map provided by \cite{Energy_Delay_Association_JSAC2011}.} \label{Fig:SumPower_Real3G_Map}
\end{figure}

Stochastic optimization theories can be exploited to solve (\ref{Eq:OriginalGreenBSAssociationResourceFormulation}) \cite{JointOptimization_HetNets_JSAC2016}. To verify the performance of our load-aware scheme, \cite{JointOptimization_HetNets_JSAC2016} compared it with the following two schemes in a real 3G BS deployment map considered in \cite{Energy_Delay_Association_JSAC2011} (cf. Fig.~3 in \cite{Energy_Delay_Association_JSAC2011}):
\begin{itemize}
  \item Fixed: The scheme that keeps all BS on and adopts fixed resource allocation regardless of traffic loads.
  \item Greedy-off: The scheme proposed in \cite{Energy_Delay_Association_JSAC2011} that optimizes BS operation and user association.
\end{itemize}
Fig.~\ref{Fig:SumPower_Real3G_Map} shows that, compared with both the fixed and Greedy-off schemes, the load-aware scheme is more energy-efficient. Moreover, the denser the BS deployment (characterized by urban, suburban, and rural areas), the more energy saving the load-aware scheme can achieve. This is because more BSs can be switched off when the network is in low traffic states.

Fig.~\ref{Fig:SumPower_Real3G_Map} exhibits great potentials of our proposed framework and techniques in energy-efficient deployment of Ud-HetNets. It is believed that the energy consumption can be further reduced when these techniques are flexibly applied to enable the network with the abilities of learning and inferring.

\section{Conclusions} \label{Section:Conlusions}
Ud-HetNets have been proposed as an important network-organization architecture to improve the network capacity. With the exponential growth of mobile traffic volumes, Ud-HetNets are bound to networks of big data. Ud-HetNets of big data impose great challenges on their practical deployment due to the incurred huge energy consumption. In this article, we have first developed a big-data-aware AI-based network framework for energy-efficient operations of Ud-HetNets. We have then identified four promising techniques, namely big data analysis, adaptive BS operation, proactive caching, and interference-aware resource allocation. The framework and these related techniques enable the network with the abilities of learning and inferring by analyzing the collected big data, and then save energy from both large scales (BS operation) and small scales (proactive caching and interference-aware resource allocation). We have further established a load-aware stochastic optimization model to show the potential of our proposed framework and techniques in reducing energy consumption. However, it is worthwhile to note that we are still at a very primary stage in these studies, and thus further investigations on data science, caching, and resource optimization are eagerly deserved for green Ud-HetNets.

\section{Acknowledgement}
This work was supported in part by National Science Foundation of China with Grants 61601192, 61631015, 61501356, 61428104, and 61471163, Joint Specialized Research Fund for the Doctoral Program of Higher Education (SRFDP) and Research Grants Council Earmarked Research Grants (RGC ERG) with Grant 20130142140002, and the Fundamental Research Funds for the Central Universities with Grant 2016YXMS298.

\bibliographystyle{IEEEtran}
\bibliography{IEEEabrv,MyRef}

\section*{Biographies}

{\footnotesize{\noindent Yuzhou~Li [M'14] (yuzhouli@hust.edu.cn) received the Ph.D. degree in communications and information systems from the School of Telecommunications Engineering, Xidian University, Xi'an, China, in December 2015. Since then, he has been with the School of Electronic Information and Communications, Huazhong University of Science and Technology, Wuhan, China, where he is currently an Assistant Professor. His research interests include 5G wireless networks, marine object detection and recognition, and undersea localization.}}

\vspace{1em}

{\footnotesize{\noindent Yu~Zhang (yu$\_$zhang@hust.edu.cn) received the B.Eng. degree in communications engineering from the School
of Electronic Information and Communications, Huazhong University of Science and Technology, Wuhan, China, where he is currently pursuing the M.S. degree. His research interests include green communications and resource allocation in wireless communications.}}

\vspace{1em}

{\footnotesize{\noindent Kai~Luo (kluo@hust.edu.cn) received his BEng degree from School of Electronics Information and Communications (EIC), Huazhong University of Science and Technology (HUST), China in 2006. Then, he received his Ph.D. degree in electrical engineering from Imperial College London in 2013. In 2013, he joined Institute of Electronics, Chinese Academy of Sciences. Since 2014, he is an Assistant Professor with School of EIC, HUST. His research interests are signal processing and MIMO communications.}}

\vspace{1em}

{\footnotesize{\noindent Tao~Jiang [M'06-SM'10] (taojiang@hust.edu.cn) is currently a Distinguished Professor with the School of Electronics Information and Communications, Huazhong University of Science and Technology, Wuhan, China. He has authored or co-authored over 200 technical papers and 5 books in the areas of wireless communications and networks. He is the associate editor-in-chief of China Communications and on the Editorial Board of IEEE Transactions on Signal Processing and on Vehicular Technology, among others.}}

\vspace{1em}

{\footnotesize{\noindent Zan~Li [M'05-SM'14] (zanli@xidian.edu.cn) received the B.S. degree in communications engineering and the M.S. and Ph.D. degrees in communication and information systems from Xidian University, Xi'an, China, in 1998, 2001, and 2004, respectively. She is currently a Professor of the State Key Laboratory of Integrated Services Networks (ISN), Xidian University. Her research interests include wireless communications and signal processing, weak signal detection, spectrum sensing, and co-operative communication.}}

\vspace{1em}

{\footnotesize{\noindent Wei~Peng [M'07-SM'12] (pengwei@hust.edu.cn) received her Ph. D. degree from the University of Hong Kong in 2007. She worked at Tohoku University, Japan, as an assistant professor from 2009 to 2013. Since 2013, she has been with the School of Electronic Information
and Communications, Huazhong University of Science and Technology, China, as an associate professor. Her research interests include signal processing for large-scale wireless communication systems, online optimization, and so on.}}

\end{document}